%% file: lepto-proc.tex
\documentclass[graybox]{svmult}

\usepackage{mathptmx}       
\usepackage{helvet}         
\usepackage{courier}        
\usepackage{type1cm}        
\usepackage{makeidx}         
\usepackage{graphicx}        
\usepackage{multicol}        
\usepackage[bottom]{footmisc}

\usepackage{amsmath, amssymb, cite, bm}

\newcommand{\mat}[1]{\bm{#1}}

\makeindex             

\begin{document}
\title*{TeV Scale Leptogenesis}
\author{P. S. Bhupal Dev}
\institute{Consortium for Fundamental Physics, School of Physics and Astronomy, University of Manchester, Manchester M13 9PL, United Kingdom, \email{bhupal.dev@manchester.ac.uk}}
%
%
\maketitle
\abstract{This is a mini-review on the mechanism of leptogenesis, with a special emphasis on low-scale leptogenesis models which are testable in foreseeable laboratory experiments at Energy and Intensity frontiers. We also stress the importance of flavor effects in the calculation of the lepton asymmetry and the necessity of a flavor-covariant framework to consistently capture these effects. } 

\section{Introduction}
\label{sec:1}
Our Universe seems to be populated exclusively with matter and essentially no antimatter. Although this asymmetry is maximal today, at high temperatures ($T\gtrsim 1$ GeV) when quark-antiquark pairs were abundant in the thermal plasma, the baryon asymmetry observed today corresponds to a tiny asymmetry at recombination~\cite{Ade:2015xua}:
\begin{align}
\eta_B \ \equiv \ \frac{n_B-n_{\overline{B}}}{n_\gamma} \ = \ \left(6.105^{+0.086}_{-0.081}\right)\times 10^{-10} \; , \label{baryo}
\end{align} 
where $n_{B(\overline{B})}$ is the number density of baryons (antibaryons) and $n_\gamma=2T^3 \zeta(3)/\pi^2$ is the number density of photons, $\zeta(x)$ being the Riemann zeta function, with $\zeta(3)\approx 1.20206$. {\em Baryogenesis} is the mechanism by which the observed baryon asymmetry of the Universe (BAU) given by Eq.~\eqref{baryo} can arise dynamically from an initially baryon symmetric phase of the Universe, or even irrespective of any initial asymmetry. 
This necessarily requires the fulfillment of three basic Sakharov conditions~\cite{Sakharov:1967dj}: (i)  baryon number ($B$) violation, which is essential for the Universe to evolve from a state with net baryon number $B=0$ to a state with $B\neq 0$; (ii) $C$ and $CP$ violation, which allow particles and anti-particles to evolve differently so that we can have an asymmetry between them; (iii) departure from thermal equilibrium, which ensures that the asymmetry does not get erased completely. The Standard Model (SM) has, in principle, all these basic ingredients, namely, (i) the triangle anomaly violates $B$ through a non-perturbative instanton effect~\cite{tHooft:1976up}, which leads to effective ($B+L$)-violating sphaleron transitions for $T\gtrsim 100$ GeV~\cite{Kuzmin:1985mm}; (ii) there is maximal $C$ violation in weak interactions and $CP$ violation due to the Kobayashi-Maswaka phase in the quark sector~\cite{Agashe:2014kda}; (iii) the departure from thermal equilibrium can be realized at the electroweak phase transition (EWPT) if it is sufficiently first order~\cite{Cohen:1990it}. However, the SM $CP$ violation turns out to be too small to account for the observed BAU~\cite{Gavela:1993ts}. In addition, for the observed value of the Higgs mass, $m_H=(125.09\pm 0.24)$ GeV~\cite{Aad:2015zhl}, the EWPT is not first order, but only a smooth crossover~\cite{Kajantie:1996mn}. Therefore, {\em the observed BAU provides a strong evidence for the existence of new physics beyond the SM.} 

Many interesting scenarios for successful baryogenesis have been proposed in beyond SM theories; see e.g.~\cite{Cline:2006ts}. Here we will focus on the mechanism of {\em leptogenesis}~\cite{Fukugita:1986hr}, which is an elegant framework to explain the BAU, while connecting it to another seemingly disparate piece of evidence for new physics beyond the SM, namely, non-zero neutrino masses; for reviews on leptogenesis, see e.g.~\cite{Davidson:2008bu, Blanchet:2012bk}. The minimal version of leptogenesis is based on the {\em type I seesaw}  mechanism~\cite{seesaw}, which requires heavy SM gauge-singlet Majorana neutrinos $N_\alpha$ (with $\alpha=1,2,3$) to explain the observed smallness of the three active neutrino masses at tree-level.  The out-of-equilibrium decays of these heavy Majorana neutrinos in an expanding Universe create a lepton asymmetry, which  is reprocessed  into  a baryon asymmetry  through  the equilibrated    $(B+L)$-violating    electroweak    sphaleron
interactions~\cite{Kuzmin:1985mm}. 

In the original scenario of thermal leptogenesis~\cite{Fukugita:1986hr}, the heavy  Majorana neutrino  masses 
are typically close  to the  Grand Unified  Theory  (GUT) scale,  
as  suggested by  natural GUT  embedding of  the seesaw
mechanism.  In fact, for a hierarchical heavy neutrino spectrum, i.e. ($m_{N_1} \ll  m_{N_{2}} < m_{N_{3}}$), 
the  light neutrino  oscillation  data impose  a  
{\it lower} limit    on    $m_{N_1}    \gtrsim   10^9$    GeV~\cite{Davidson:2002qv}. As a consequence, 
such `vanilla' leptogenesis scenarios~\cite{Buchmuller:2004nz} are very 
difficult  to  test  in  any foreseeable experiment. Moreover, these high-scale thermal  leptogenesis scenarios run into difficulties,  
when   embedded  within  supergravity   models  of   inflation. In particular, it leads to  a potential conflict with an {\em upper}  bound on the reheat 
temperature of the Universe, $T_R  \lesssim 10^6$--$10^9$ GeV, as required to avoid overproduction of  gravitinos whose late decays may otherwise
ruin   the      success     of      Big     Bang
Nucleosynthesis~\cite{Khlopov:1984pf}.  

An attractive scenario that avoids the aforementioned problems is {\em resonant   leptogenesis} (RL)~\cite{Pilaftsis:1997dr, Pilaftsis:2003gt}, where  the $\varepsilon$-type $CP$ asymmetries due to the self-energy
effects~\cite{Flanz:1994yx, Covi:1996wh, Buchmuller:1997yu} in the heavy  Majorana neutrino decays get resonantly enhanced. This happens when the masses of at least two of the heavy neutrinos become quasi-degenerate, with a mass difference comparable to their  decay  widths~\cite{Pilaftsis:1997dr}.  The resonant enhancement of the $CP$ asymmetry allows one to avoid the lower bound on $m_{N_1}    \gtrsim   10^9$    GeV~\cite{Davidson:2002qv} and have successful leptogenesis at an experimentally accessible energy scale~\cite{Pilaftsis:2003gt, Pilaftsis:2005rv},  while retaining perfect agreement with the light-neutrino oscillation data. The level of testability is further extended in the scenario of Resonant $l$-Genesis (RL$_{l}$), where the final lepton asymmetry is dominantly generated and stored in a {\it single} lepton flavor $l$~\cite{Pilaftsis:2004xx, Deppisch:2010fr, Dev:2014laa, Dev:2015wpa}. In such models, the heavy neutrinos could be as light as the electroweak scale, while still having sizable couplings to other charged-lepton flavors $l'\neq l$, thus giving rise to potentially large lepton flavor violation (LFV) effects. 
In this mini-review, we will mainly focus on low-scale leptogenesis scenarios, which may be directly tested at the Energy~\cite{Deppisch:2015} and Intensity~\cite{Alekhin:2015byh}  frontiers. For brevity, we will only discuss the type I seesaw-based leptogenesis models; for other leptogenesis scenarios, see e.g.~\cite{Davidson:2008bu, Hambye:2012fh}.  

\section{Basic Picture}
Our starting point is the minimal type I seesaw extension of the SM Lagrangian:
\begin{align}
{\cal L} \ = \ {\cal L}_{\rm SM}+ i\overline{N}_{{\rm R}, \alpha}\gamma_\mu \partial ^\mu N_{{\rm R}, \alpha} - h_{l\alpha}\overline{L}_l\widetilde{\Phi}N_{{\rm R}, \alpha} - \frac{1}{2}\overline{N}^C_{{\rm R}, \alpha}(M_{N})_{\alpha\beta}N_{{\rm R}, \beta} +{\rm H.c.} \; ,
\end{align}
where $N_{{\rm R},\alpha}  \equiv \frac{1}{2}(\mathbf{1}+\gamma_5) N_\alpha$ are  the 
right-handed (RH) heavy  neutrino  fields, $L_l\equiv (\nu_l ~~ l)_L^{\sf T}$ are the $SU(2)_L$ lepton doublets (with $l=e,\mu,\tau$) and $\widetilde{\Phi}\equiv  i\sigma_2\Phi^*$, $\Phi$ being the SM Higgs doublet and $\sigma_2$ being the second Pauli matrix. The complex Yukawa couplings $h_{l\alpha}$ induce $CP$-violating decays of the heavy Majorana neutrinos, if kinematically allowed: $N_\alpha\to L_l\Phi$ with a decay rate $\Gamma_{l\alpha}$ and the $CP$-conjugate process $N_\alpha \to L_l^c\Phi^c$ with a decay rate $\Gamma^c_{l\alpha}$ (the shorthand notation $c$ denotes $CP$). The flavor-dependent $CP$ asymmetry can be defined as 
\begin{align}
 \varepsilon_{l\alpha} \ = \ \frac{\Gamma_{l\alpha} 
    -  \Gamma^c_{l\alpha}}
  {\sum_{k}\big(\Gamma_{k\alpha} 
      +  \Gamma^{c}_{k\alpha}\big)} \
  \equiv \ \frac{\Delta \Gamma_{l\alpha}}{\Gamma_{N_\alpha}}\;,
  \label{eps}
\end{align} 
where  $\Gamma_{N_\alpha}$  is the  total  decay  width  of the  heavy
Majorana neutrino $N_\alpha$ which, at tree-level, is given by $\Gamma_{N_\alpha}=\frac{m_{N_{\alpha}}}{8\pi}(h^\dag h)_{\alpha\alpha}$. 
%
A non-zero $CP$ asymmetry arises at one-loop level due to the interference between the tree-level graph with either the vertex or the self-energy graph. Following the terminology used for $CP$ violation in neutral meson systems, we denote these two contributions as $\varepsilon'$-type and $\varepsilon$-type $CP$ violation respectively. In the two heavy-neutrino case ($\alpha,\beta=1,2$; $\alpha\neq \beta$), they can be expressed in a simple analytic form~\cite{Covi:1996wh, Pilaftsis:2003gt}:  
\begin{eqnarray}
 \varepsilon'_{l\alpha} \ &=& \ \frac{{\rm Im}
    \left[({h}^*_{l\alpha} {h}_{l\beta})({h}^\dag {h})_{\alpha\beta}\right]}
  {8\pi \: ({h}^\dag {h})_{\alpha\alpha}} \frac{m_{N_\beta}}{m_{N_\alpha}}\left[1-\bigg(1+\frac{m^2_{N_\beta}}{m^2_{N_\alpha}}\bigg)\ln\bigg(1+ \frac{m^2_{N_\alpha}}{m^2_{N_\beta}}\bigg)\right]
 \; , \label{epsp}\\
\varepsilon_{l\alpha} \ &=& \ \frac{{\rm Im}\left[(h^*_{l\alpha} {h}_{l\beta})({h}^\dag {h})_{\alpha\beta}\right]+\frac{m_{N_\alpha}}{m_{N_\beta}}\:{\rm Im}\left[({h}^*_{l\alpha} {h}_{l\beta})({h}^\dag {h})_{\beta\alpha}\right]}
  {8\pi \: ({h}^\dag {h})_{\alpha\alpha}} \frac{(m^2_{N_\alpha}-m^2_{N_\beta})
    m_{N_\alpha}m_{N_\beta}}
  {(m^2_{N_\alpha}-m^2_{N_\beta})^2
    +m^2_{N_\alpha}\Gamma^2_{N_\beta}} \; . \nonumber \\ 
\label{eps} 
 \end{eqnarray}
In the quasi-degenerate limit $|m_{N_\alpha}-m_{N_\beta}|\sim \frac{1}{2}\Gamma_{N_{\alpha,\beta}}$, the $\varepsilon$-type contribution becomes resonantly enhanced, as is evident from the second denominator in Eq.~\eqref{eps}. 

Due to the Majorana nature of the heavy neutrinos, their decays to lepton and Higgs fields violate lepton number which, in presence of $CP$ violation, leads to the generation of a lepton (or $B-L$) asymmetry. Part of this asymmetry is washed out due to the inverse decay processes $L\Phi\to N, L^c\Phi^c \to N$ and various $\Delta L=1$ (e.g. $NL\leftrightarrow Q u^c$) and $\Delta L=2$ (e.g. $L\Phi \leftrightarrow L^c\Phi^c$) scattering processes. In the flavor-diagonal limit, the total amount of $B-L$ asymmetry generated at a given temperature can be calculated by solving the following set of coupled Boltzmann equations~\cite{Buchmuller:2004nz}:
\begin{align}
\frac{d{\cal N}_{N_\alpha}}{dz} \ & = \  -(D_\alpha+S_\alpha)({\cal N}_{N_\alpha}-{\cal N}_{N_\alpha}^{\rm eq}) \; , \label{be1}\\
\frac{d{\cal N}_{B-L}}{dz} \ & = \ \sum_{\alpha} \varepsilon_\alpha D_\alpha ({\cal N}_{N_\alpha}-{\cal N}_{N_\alpha}^{\rm eq}) - {\cal N}_{B-L} \sum_\alpha W_\alpha \; , \label{be2}
\end{align}
where $z\equiv m_{N_1}/T$, with $N_1$ being the lightest heavy neutrino, ${\cal N}_X$ denotes the number density of $X$ in a portion of co-moving volume containing one heavy-neutrino in ultra-relativistic limit, so that ${\cal N}^{\rm eq}_{N_\alpha}(T\gg m_{N_\alpha})=1$, $\varepsilon_\alpha\equiv \sum_l(\varepsilon_{l\alpha}+\varepsilon'_{l\alpha})$ is the total $CP$ asymmetry due to the decay of $N_\alpha$,  and $D_\alpha, S_\alpha, W_\alpha$ denote the decay, scattering and washout rates, respectively. Given the Hubble expansion rate $H(T)\simeq 1.66g_*\frac{T^2}{M_{\rm Pl}}$, where $g_*$ is the total relativistic degrees of freedom and $M_{\rm Pl}=1.22\times 10^{19}$ GeV is the Planck mass, we define the decay parameters $K_\alpha \equiv \frac{\Gamma_{D_\alpha}(T=0)}{H(T=m_{N_\alpha})}$, where $\Gamma_{D_\alpha}\equiv \sum_l(\Gamma_{l\alpha}+\Gamma^c_{l\alpha})$. For $K_\alpha\gg 1$, the system is in the {\em strong washout} regime, where the final lepton asymmetry is insensitive to any initial asymmetry present. The decay rates are given by $D_\alpha \equiv \frac{\Gamma_{D_\alpha}}{Hz} = K_\alpha x_\alpha z \frac{{\cal K}_1(z)}{{\cal K}_2(z)}$, where $x_\alpha \equiv \frac{m^2_{N_\alpha}}{m^2_{N_1}}$ and ${\cal K}_n(z)$ is the $n$th-order modified Bessel function of the second kind. Similarly, the washout rates induced by inverse decays are given by $W_\alpha^{\rm ID} = \frac{1}{4}K_\alpha\sqrt{x_\alpha} {\cal K}_1(z_\alpha)z_\alpha^3$, where $z_\alpha\equiv z\sqrt{x_\alpha}$. Other washout terms due to $2\leftrightarrow 2$ scattering can be similarly calculated~\cite{Buchmuller:2004nz}. The final $B-L$ asymmetry is given by 
${\cal N}_{B-L}^{\rm f}  =  \sum_\alpha \varepsilon_\alpha \kappa_\alpha(z\to \infty)$, 
where $\kappa_\alpha(z)$'s are the efficiency factors, obtained from Eqs.~\eqref{be1} and \eqref{be2}: 
\begin{align}
\kappa_\alpha(z) \ = \ -\int_{z_{\rm in}}^z dz'\: \frac{D_\alpha(z')}{D_\alpha(z')+S_\alpha(z')}\: \frac{d{\cal N}_{N_\alpha}}{dz'} \: \exp\bigg[-\int_{z'}^z dz''\sum_\alpha W_\alpha(z'')\bigg] \; .
\end{align}        

At temperatures $T\gg 100$ GeV, when the $(B+L)$-violating electroweak sphalerons are in thermal equilibrium, a fraction $a_{\rm sph}=\frac{28}{79}$ of the $B-L$ asymmetry is reprocessed to a baryon asymmetry~\cite{Khlebnikov:1988sr}. There is an additional entropy dilution factor $f=\frac{{\cal N}_\gamma^{\rm rec}}{{\cal N}_\gamma,*} = \frac{2387}{86}$ due to the standard photon production from the onset of leptogenesis till the epoch of recombination~\cite{Kolb:1990vq}. Combining all these effects, the predicted baryon asymmetry due to the mechanism of leptogenesis is given by 
\begin{align}
\eta_B \ = \ \frac{a_{\rm sph}}{f}{\cal N}^{\rm f}_{B-L} \ \simeq \ 10^{-2} \: \sum_\alpha \varepsilon_\alpha \kappa_\alpha(z\to \infty) \; ,
\end{align}  
which has to be compared with the observed BAU given by Eq.~\eqref{baryo}. 

\section{Flavor Effects}
Flavor effects in both heavy-neutrino and charged-lepton sectors, as well as the interplay between them, can play an important role in determining the final lepton asymmetry, especially in low-scale leptogenesis models~\cite{Blanchet:2012bk}. These intrinsically-quantum effects  can, in principle, be accounted for by extending the flavor-diagonal Boltzmann equations \eqref{be1} and \eqref{be2} for the number  densities of individual flavor species to a  semi-classical evolution equation for a {\it matrix  of  number densities}, analogous to the formalism presented in~\cite{Sigl:1993} for light neutrinos. This so-called `density matrix' formalism has been adopted to describe flavor effects in various leptogenesis scenarios~\cite{Abada:2006fw, Nardi:2006fx, Akhmedov:1998qx}. It was recently shown~\cite{Dev:2014laa, Dev:2014tpa}, in a semi-classical approach, that a consistent treatment of {\em all} pertinent flavor effects, including flavor mixing, oscillations and off-diagonal (de)coherences, necessitates a {\em fully} flavor-covariant formalism. It was further shown that the resonant mixing of different heavy-neutrino flavors and coherent oscillations between them are two {\em distinct} physical phenomena, whose contributions to the $CP$ asymmetry could be of similar order of magnitude in the resonant regime. Note that this is analogous to the experimentally-distinguishable phenomena of mixing and oscillations in the neutral meson systems~\cite{Agashe:2014kda}.

One can go beyond the semi-classical `density-matrix' approach to leptogenesis by means of a quantum field-theoretic analogue of the Boltzmann equations, known as the Kadanoff-Baym (KB) equations~\cite{Baym:1961zz}. Such `first-principles' approaches to leptogenesis~\cite{Buchmuller:2000nd} are, in principle, capable of accounting consistently for all flavor effects, in addition to off-shell and finite-width effects, including thermal corrections. However, it is often necessary to use truncated gradient expansions and quasi-particle ansaetze to relate the propagators appearing in the KB equations to particle number densities. Recently, using a perturbative formulation of thermal field theory~\cite{Millington:2012pf}, it was shown~\cite{Dev:2014wsa} that quantum transport equations for leptogenesis can be obtained from the KB formalism without the need for gradient expansion or quasi-particle ansaetze, thereby capturing fully the pertinent flavor effects.
Specifically, the source term for the lepton asymmetry obtained, at leading order, in this KB approach~\cite{Dev:2014wsa} was found to be exactly the same as that obtained in the semi-classical flavor-covariant approach of~\cite{Dev:2014laa}, confirming that flavor mixing and oscillations are indeed two {\em physically-distinct} phenomena, at least in the weakly-resonant regime. The proper treatment of these flavor effects can lead to a significant enhancement of the final lepton asymmetry, as compared to partially flavor-dependent or flavor-diagonal limits~\cite{Dev:2014laa, Dev:2015wpa}, thereby enlarging the viable parameter space for models of RL and enhancing the prospects of testing the leptogenesis mechanism. 

\section{Phenomenology} 
As an example of a testable scenario of leptogenesis, we consider a minimal $\mathrm{RL}_l$ model that possesses an approximate SO(3)-symmetric heavy-neutrino sector at some high scale $\mu_X$, with mass matrix $\bm{M}_N(\mu_X)=m_N\bm{1}_3+\bm{\Delta M}_N$~\cite{Pilaftsis:2005rv, Deppisch:2010fr}, where the SO(3)-breaking mass term is of the  form $\bm{\Delta M}_N=\mathrm{diag}(\Delta M_1,\Delta M_2/2,-\Delta M_2/2)$~\cite{Dev:2015wpa}. By virtue of the renormalization group  running, an additional mass splitting term 
\begin{equation}
 \mat{\Delta  M}_N^{\rm RG} \ \simeq \ - \,\frac{m_N}{8\pi^2}
  \ln\left(\frac{\mu_X}{m_N}\right)
  {\rm Re}\left[\mat{h}^\dag(\mu_X) \mat{h}(\mu_X)\right] 
  \label{deltam}
\end{equation}
 is induced at the scale relevant to RL. In order to ensure the smallness of the light-neutrino masses, we also require the heavy-neutrino Yukawa sector to have an approximate leptonic U(1)$_l$ symmetry. As an explicit example, we consider an RL$_\tau$ model, with the following Yukawa coupling structure~\cite{Pilaftsis:2004xx,Pilaftsis:2005rv}:
\begin{eqnarray}
  \bm{h} \ = \ \left(\begin{array}{ccc}
      \epsilon_e & a \,e^{-i\pi/4} & a\,e^{i\pi/4}\\
      \epsilon_\mu & b\,e^{-i\pi/4} & b\,e^{i\pi/4}\\
      \epsilon_\tau & c\,e^{-i\pi/4} & c\,e^{i\pi/4}
    \end{array}\right)  \; , 
  \label{yuk}
\end{eqnarray}
where $a,b,c$ are arbitrary complex parameters and $\epsilon_{e,\mu,\tau}$ are the perturbation terms that break the $U(1)$ symmetry. In order to be consistent with the observed neutrino-oscillation data, we require $|a|,|b|\lesssim 10^{-2}$ for electroweak scale heavy neutrinos. In addition, in order to protect the $\tau$ asymmetry from large washout effects, we require $|c|\lesssim 10^{-5}\ll|a|,|b|$ and $|\epsilon_{e,\mu,\tau}|\lesssim 10^{-6}$. 

A choice of benchmark values for these parameters, satisfying all the current experimental constraints and allowing successful leptogenesis, is given below: 
\begin{align}
& m_N \ = \ 400~{\rm GeV}, \quad    
\frac{\Delta M_1}{m_N} \ = \ -3\times 10^{-5}, \quad  
\frac{\Delta M_2}{m_N} \ = \ (-1.21+0.10\,i)\times 10^{-9}, \nonumber \\ 
& a \ = \ (4.93-2.32 \, i)\times 10^{-3}, \quad 
      b \ = \ (8.04 - 3.79 \, i)\times 10^{-3}, \quad 
 c \ =\ 2\times 10^{-7}, \nonumber \\ 
  &    \epsilon_e \ = \ 5.73\, {i}\times 10^{-8}, \quad 
      \epsilon_\mu \ =\ 4.30\, {i}\times 10^{-7}, \quad 
      \epsilon_\tau \ = \ 6.39\, {i}\times 10^{-7} .
\end{align}
The corresponding predictions for some low-energy LFV observables are given by 
\begin{align}
& {\rm BR}(\mu\to e\gamma) \ = \ 1.9\times 10^{-13}, \quad 
{\rm BR}(\mu^- \to e^-e^+e^-) \ =\ 9.3\times 10^{-15}, \nonumber \\
& R_{\mu\to e}^{\text{Ti}} \ = \ 2.9\times 10^{-13}, \quad 
R_{\mu\to e}^{\text{Au}} \ = \ 3.2\times 10^{-13}, \quad 
R_{\mu\to e}^{\text{Pb}} \ = \ 2.2\times 10^{-13},
\end{align}
all of which can be probed in future at the Intensity frontier~\cite{Hewett:2014qja}. Similarly, sub-TeV heavy Majorana neutrinos with ${\cal O}(10^{-2})$ Yukawa couplings are directly accessible in the run-II phase of the LHC~\cite{Dev:2013wba} as well as at future lepton colliders~\cite{Banerjee:2015gca}. 

In general, any observation of lepton number violation (LNV) at the LHC will yield a lower bound on the washout factor for the lepton asymmetry and could falsify {\em high}-scale leptogenesis as a viable mechanism behind the observed BAU~\cite{Deppisch:2013jxa}. However, one should keep in mind possible exceptions to this general argument, e.g. scenarios where LNV is confined to a specific flavor sector, models with new symmetries and/or conserved charges which could stabilize the baryon asymmetry against LNV washout, and models where lepton asymmetry can be generated below the observed LNV scale. An important related question is whether {\em low}-scale leptogenesis models could be ruled out from experiments. This has been investigated~\cite{Frere:2008ct, Dev:2014iva, Dev:2015vra, Dhuria:2015cfa} in the context of Left-Right symmetric models 
and it was shown that the minimum value of the RH gauge boson mass for successful leptogenesis, while satisfying all experimental constraints in the low-energy sector, is about 10 TeV~\cite{Dev:2015vra}. Thus, any positive signal for an RH gauge boson at the LHC might provide a litmus test for the mechanism of leptogenesis.

\section{Conclusion}
Leptogenesis is an attractive mechanism for dynamically generating the observed baryon asymmetry of the Universe, while relating it to the origin of neutrino mass.  Resonant leptogenesis allows the relevant energy scale to be as low as the electroweak scale, thus offering a unique opportunity to test this idea in laboratory experiments. Flavor effects play an important role in the predictions for the lepton asymmetry, and hence, for the testability of the low-scale leptogenesis models. We have illustrated that models of resonant leptogenesis could lead to observable effects in current and future experiments, and may even be falsified in certain cases.

\begin{acknowledgement}
I thank the organizers of the XXI DAE-BRNS HEP Symposium for the invitation and IIT, Guwahati  for the local hospitality. 
This work was supported by the Lancaster-Manchester-Sheffield Consortium for Fundamental
 Physics under STFC grant ST/L000520/1. 
\end{acknowledgement}
%

\input{lepto-ref}

\end{document}

%% file: lepto-ref.tex
%
%

%% file: lepto-proc.bbl
\begin{thebibliography}{99.}%

\bibitem{Ade:2015xua} 
  P.~A.~R.~Ade {\it et al.}  [Planck Collaboration],
  arXiv:1502.01589 [astro-ph.CO].

\bibitem{Sakharov:1967dj} 
  A.~D.~Sakharov,
  JETP Lett.\  {\bf 5}, 24 (1967). 

\bibitem{tHooft:1976up} 
  G.~'t Hooft,
  Phys.\ Rev.\ Lett.\  {\bf 37}, 8 (1976).

\bibitem{Kuzmin:1985mm} 
  V.~A.~Kuzmin, V.~A.~Rubakov and M.~E.~Shaposhnikov,
  Phys.\ Lett.\ B {\bf 155}, 36 (1985).

\bibitem{Agashe:2014kda} 
  K.~A.~Olive {\it et al.}  [Particle Data Group Collaboration],
  Chin.\ Phys.\ C {\bf 38}, 090001 (2014).


\bibitem{Cohen:1990it} 
  A.~G.~Cohen, D.~B.~Kaplan and A.~E.~Nelson,
  Nucl.\ Phys.\ B {\bf 349}, 727 (1991).  

\bibitem{Gavela:1993ts} 
  M.~B.~Gavela, P.~Hernandez, J.~Orloff and O.~Pene,
  Mod.\ Phys.\ Lett.\ A {\bf 9}, 795 (1994). 

\bibitem{Aad:2015zhl} 
  G.~Aad {\it et al.}  [ATLAS and CMS Collaborations],
  Phys.\ Rev.\ Lett.\  {\bf 114}, 191803 (2015). 

\bibitem{Kajantie:1996mn} 
  K.~Kajantie, M.~Laine, K.~Rummukainen and M.~E.~Shaposhnikov,
  Phys.\ Rev.\ Lett.\  {\bf 77}, 2887 (1996); 
  K.~Rummukainen, M.~Tsypin, K.~Kajantie, M.~Laine and M.~E.~Shaposhnikov,
  Nucl.\ Phys.\ B {\bf 532}, 283 (1998). 

\bibitem{Cline:2006ts} 
  J.~M.~Cline, {\em Les Houches Lectures}, 
  hep-ph/0609145.

\bibitem{Fukugita:1986hr} 
  M.~Fukugita and T.~Yanagida,
  Phys.\ Lett.\ B {\bf 174}, 45 (1986).

\bibitem{Davidson:2008bu} 
  S.~Davidson, E.~Nardi and Y.~Nir,
  Phys.\ Rept.\  {\bf 466}, 105 (2008). 

\bibitem{Blanchet:2012bk} 
  S.~Blanchet and P.~Di Bari,
  New J.\ Phys.\  {\bf 14}, 125012 (2012). 

\bibitem{seesaw}
  P. Minkowski,
  Phys. Lett. B {\bf 67}, 421 (1977); 
%
  R. N. Mohapatra and G. Senjanovi\'{c}, 
  Phys. Rev. Lett. {\bf 44}, 912 (1980); 
%
  M. Gell-Mann, P. Ramond and R. Slansky, 
  Conf.\ Proc.\ C {\bf 790927}, 315 (1979); 
%
  T. Yanagida,
  in {\it Proceedings of the Workshop on Unified Theories
    and Baryon Number in the Universe},
  eds. A. Sawada and A. Sugamoto, KEK, Tsukuba (1979); 
%
J.~Schechter and J.~W.~F.~Valle,
  Phys.\ Rev.\ D {\bf 22}, 2227 (1980).


\bibitem{Davidson:2002qv} 
  S.~Davidson and A.~Ibarra,
  Phys.\ Lett.\ B {\bf 535}, 25 (2002); 
%
  W.~Buchm\"{u}ller, P.~Di Bari and M.~Pl\"{u}macher,
  Nucl.\ Phys.\ B {\bf 643}, 367 (2002)
  [Erratum-ibid.\ B {\bf 793}, 362 (2008)]; 
%
  T.~Hambye, Y.~Lin, A.~Notari, M.~Papucci and A.~Strumia,
  Nucl.\ Phys.\ B {\bf 695}, 169 (2004). 

\bibitem{Buchmuller:2004nz} 
  W.~Buchm\"{u}ller, P.~Di Bari and M.~Pl\"{u}macher,
  Annals Phys.\  {\bf 315}, 305 (2005). 

\bibitem{Khlopov:1984pf} 
  M.~Yu.~Khlopov and A.~D.~Linde,
  Phys.\ Lett.\ B {\bf 138}, 265 (1984);
%
  J.~R.~Ellis, J.~E.~Kim and D.~V.~Nanopoulos,
  Phys.\ Lett.\ B {\bf 145}, 181 (1984); 
%
%
%
%
%
  M.~Kawasaki, K.~Kohri, T.~Moroi and A.~Yotsuyanagi,
  Phys.\ Rev.\ D {\bf 78}, 065011 (2008). 

\bibitem{Pilaftsis:1997dr} 
  A.~Pilaftsis,
  Nucl.\ Phys.\ B {\bf 504}, 61 (1997); 
%
  Phys.\ Rev.\ D {\bf 56}, 5431 (1997). 

\bibitem{Pilaftsis:2003gt} 
A.~Pilaftsis and T.~E.~J.~Underwood,
  Nucl.\ Phys.\ B {\bf 692}, 303 (2004). 

%
\bibitem{Flanz:1994yx} 
  M.~Flanz, E.~A.~Paschos and U.~Sarkar,
  Phys.\ Lett.\ B {\bf 345}, 248 (1995)
  [Erratum-ibid.\ B {\bf 382}, 447 (1996)]. 

\bibitem{Covi:1996wh} 
  L.~Covi, E.~Roulet and F.~Vissani,
  Phys.\ Lett.\ B {\bf 384}, 169 (1996). 

\bibitem{Buchmuller:1997yu} 
  W.~Buchmuller and M.~Plumacher,
  Phys.\ Lett.\ B {\bf 431}, 354 (1998). 

\bibitem{Pilaftsis:2005rv} 
  A.~Pilaftsis and T.~E.~J.~Underwood,
  Phys.\ Rev.\ D {\bf 72}, 113001 (2005). 

\bibitem{Pilaftsis:2004xx} 
  A.~Pilaftsis,
  Phys.\ Rev.\ Lett.\  {\bf 95}, 081602 (2005). 

\bibitem{Deppisch:2010fr} 
F.~F.~Deppisch and A.~Pilaftsis,
  Phys.\ Rev.\ D {\bf 83}, 076007 (2011). 

\bibitem{Dev:2014laa}
  P.~S.~B.~Dev, P.~Millington, A.~Pilaftsis and D.~Teresi,
 Nucl.\ Phys.\ B {\bf 886}, 569 (2014).

\bibitem{Dev:2015wpa} 
  P.~S.~B.~Dev, P.~Millington, A.~Pilaftsis and D.~Teresi,
  arXiv:1504.07640 [hep-ph].

\bibitem{Deppisch:2015} F.~F.~Deppisch, P.~S.~B.~Dev and A.~Pilaftsis, arXiv:1502.06541 [hep-ph].

\bibitem{Alekhin:2015byh} 
  S.~Alekhin {\it et al.},
  arXiv:1504.04855 [hep-ph].

\bibitem{Hambye:2012fh} 
  T.~Hambye,
  New J.\ Phys.\  {\bf 14}, 125014 (2012). 

\bibitem{Khlebnikov:1988sr} 
  S.~Y.~Khlebnikov and M.~E.~Shaposhnikov,
  Nucl.\ Phys.\ B {\bf 308}, 885 (1988); 
  J.~A.~Harvey and M.~S.~Turner,
  Phys.\ Rev.\ D {\bf 42}, 3344 (1990).

\bibitem{Kolb:1990vq} 
  E.~W.~Kolb and M.~S.~Turner,
  Front.\ Phys.\  {\bf 69}, 1 (1990).

\bibitem{Sigl:1993} 
  G.~Sigl and G.~Raffelt,
  Nucl.\ Phys.\ B {\bf 406}, 423 (1993).

\bibitem{Abada:2006fw} 
  A.~Abada, S.~Davidson, F.~-X.~Josse-Michaux, M.~Losada and A.~Riotto,
  JCAP {\bf 0604}, 004 (2006);
  A.~Abada, S.~Davidson, A.~Ibarra, F.~-X.~Josse-Michaux,
  M.~Losada and A.~Riotto,
  JHEP {\bf 0609}, 010 (2006).

\bibitem{Nardi:2006fx} 
  E.~Nardi, Y.~Nir, E.~Roulet and J.~Racker,
  JHEP {\bf 0601}, 164 (2006);
  S.~Blanchet and P.~Di Bari,
  JCAP {\bf 0703}, 018 (2007); 
  A.~De Simone and A.~Riotto,
  JCAP {\bf 0702}, 005 (2007);
  S.~Blanchet, P.~Di Bari, D.~A.~Jones and L.~Marzola,
  JCAP {\bf 1301}, 041 (2013). 

\bibitem{Akhmedov:1998qx} 
  E.~K.~Akhmedov, V.~A.~Rubakov and A.~Y.~Smirnov,
  Phys.\ Rev.\ Lett.\  {\bf 81}, 1359 (1998); 
%
  T.~Asaka and M.~Shaposhnikov,
  Phys.\ Lett.\ B {\bf 620}, 17 (2005); 
%
%
  J.~S.~Gagnon and M.~Shaposhnikov,
  Phys.\ Rev.\ D {\bf 83}, 065021 (2011); 
%
  T.~Asaka, S.~Eijima and H.~Ishida,
  JCAP {\bf 1202}, 021 (2012); 
%
  L.~Canetti, M.~Drewes, T.~Frossard and M.~Shaposhnikov,
  Phys.\ Rev.\ D {\bf 87}, 093006 (2013); 
%
  B.~Shuve and I.~Yavin,
  Phys.\ Rev.\ D {\bf 89}, 075014 (2014). 

\bibitem{Dev:2014tpa} 
  P.~S.~B. Dev, P.~Millington, A.~Pilaftsis and D.~Teresi,
  arXiv:1409.8263 [hep-ph]; 
  arXiv:1502.07987 [hep-ph].

\bibitem{Baym:1961zz}
  L. Kadanoff and G. Baym,
  {\it Quantum Statistical Mechanics}, Benjamin, New York (1962).


\bibitem{Buchmuller:2000nd} 
  W.~Buchm\"{u}ller and S.~Fredenhagen,
  Phys.\ Lett.\ B {\bf 483}, 217 (2000); 
%
  A.~De Simone and A.~Riotto,
  JCAP {\bf 0708}, 002 (2007); 
%
%
%
%
%
  M.~Garny, A.~Hohenegger, A.~Kartavtsev and M.~Lindner,
  Phys.\ Rev.\ D 
  {\bf 81}, 085027 (2010); 
%
  V.~Cirigliano, C.~Lee, M.~J.~Ramsey-Musolf and S.~Tulin,
  Phys.\ Rev.\ D {\bf 81}, 103503 (2010); 
%
%
%
%
  M.~Beneke, B.~Garbrecht, C.~Fidler, M.~Herranen and P.~Schwaller,
  Nucl.\ Phys.\ B {\bf 843}, 177 (2011); 
%
%
  A.~Anisimov, W.~Buchm\"{u}ller, M.~Drewes and S.~Mendizabal,
  Annals Phys.\  {\bf 326}, 1998 (2011)
  [Erratum-ibid.\  {\bf 338}, 376 (2011)]; 
%
  B.~Garbrecht and M.~Herranen,
  Nucl.\ Phys.\ B {\bf 861}, 17 (2012); 
%
  M.~Garny, A.~Kartavtsev and A.~Hohenegger,
  Annals Phys.\  {\bf 328}, 26 (2013); 
%
%
%
  T.~Frossard, M.~Garny, A.~Hohenegger, A.~Kartavtsev and D.~Mitrouskas,
  Phys.\ Rev.\ D {\bf 87}, 085009 (2013); 
%
%
%
%
  S.~Iso, K.~Shimada and M.~Yamanaka,
  JHEP {\bf 1404}, 062 (2014); 
%
%
  A.~Hohenegger and A.~Kartavtsev,
  JHEP {\bf 1407}, 130 (2014). 
%

\bibitem{Millington:2012pf}
  P.~Millington and A.~Pilaftsis,
  Phys.\ Rev.\ D {\bf 88}, 085009 (2013). 
   Phys.\ Lett.\ B {\bf 724}, 56 (2013). 

\bibitem{Dev:2014wsa}
  P.~S.~B.~Dev, P.~Millington, A.~Pilaftsis and D.~Teresi,
  Nucl.\ Phys.\ B {\bf 891}, 128 (2015). 

\bibitem{Hewett:2014qja} 
  J.~L.~Hewett {\it et al.},
  arXiv:1401.6077 [hep-ex].

\bibitem{Dev:2013wba} 
  P.~S.~B.~Dev, A.~Pilaftsis and U.~K.~Yang,
  Phys.\ Rev.\ Lett.\  {\bf 112}, 081801 (2014). 

\bibitem{Banerjee:2015gca} 
  S.~Banerjee, P.~S.~B.~Dev, A.~Ibarra, T.~Mandal and M.~Mitra,
  arXiv:1503.05491 [hep-ph].

\bibitem{Deppisch:2013jxa} 
  F.~F.~Deppisch, J.~Harz and M.~Hirsch,
  Phys.\ Rev.\ Lett.\  {\bf 112}, 221601 (2014). 

\bibitem{Frere:2008ct} 
  J.~M.~Frere, T.~Hambye and G.~Vertongen,
  JHEP {\bf 0901}, 051 (2009). 

\bibitem{Dev:2014iva} 
  P.~S.~B. Dev, C.~H.~Lee and R.~N.~Mohapatra,
  Phys.\ Rev.\ D {\bf 90}, 095012 (2014). 

\bibitem{Dev:2015vra} 
  P.~S.~B. Dev, C.~H.~Lee and R.~N.~Mohapatra,
  arXiv:1503.04970 [hep-ph]. 

\bibitem{Dhuria:2015cfa} 
  M.~Dhuria, C.~Hati, R.~Rangarajan and U.~Sarkar,
  arXiv:1503.07198 [hep-ph].

\end{thebibliography}
